# A synthetic autonomous rotary nanomotor made from and fuelled by DNA

*Katherine E. Dunn*

Formerly at: Clarendon Laboratory, Department of Physics, University of Oxford, Parks Road, Oxford, OX1 3PU, UK.

Now at: Department of Electronics, University of York, Heslington, York, YO10 5DD, UK

Correspondence to K.E.D: katherine.dunn@york.ac.uk

Date: 30$^{th}$ May 2015

**Abstract**

DNA nanostructures are made using synthetic DNA strands, the sequences of which are designed such that they will self-assemble into the desired form by hybridization of complementary domains. Various structures and devices have been presented, including DNA tweezers, nanorobots and a range of linear motors such as bipedal walkers. Inspiration for the latter is drawn from naturally occurring molecular motors like kinesin.

This paper describes a concept for an autonomous rotary nanomotor made from DNA, which utilizes the well-known and widely-studied phenomenon of toehold-mediated DNA strand displacement. The motor is to be driven by a series of strand displacement reactions, the order of which is controlled by steric constraints arising from the secondary structure of the DNA strands comprising the motor mechanism.

The capabilities of DNA motors would be extended significantly if autonomous rotary motion could be achieved. The device has a range of potential applications, including molecular computation and single-molecule manipulation.

**Introduction**

Less than thirty years after the discovery of the structure of DNA, Seeman proposed that artificial nanostructures could be constructed using synthetic DNA strands with sequences designed to promote self-assembly of a particular structure through base-pairing between complementary domains [1-3]. Since then a variety of DNA nanostructures have been prepared, including the DNA cube [4], 'tiles' [5], polyhedra [6,7], two- [8] and three- [9] dimensional crystals. Recently, objects of around 100nm in size have been synthesized using the technique of DNA origami [10-12], which involves folding a long single-stranded DNA scaffold (thousands of nucleotides long) into a designed shape by means of hybridization with hundreds of short DNA strands known as staples.

Dynamic DNA devices can be constructed by using toehold-mediated strand displacement [13] to drive conformational changes. Toehold-mediated strand displacement occurs in a system consisting of three DNA strands, two of which are initially hybridized to form a duplex with a short single-stranded overhang (the 'toehold'). The base sequence in the third strand is such that it can bind to the toehold and displace the shorter of the two strands, maximizing the number of base pairs. The presence of mismatches can reduce displacement rate [14], and the effectiveness of a toehold can be decreased by up to 100-fold when it is concealed inside the loop of a DNA hairpin [15].

Examples of dynamic devices constructed from DNA include 'tweezers' [16] and 'motors' [17] such as 'walkers' [18,19] inspired by natural molecular motors like kinesin. Previous rotary motors made from DNA include a tile which changed conformation by rotating upon addition of 'set' or 'unset' strands [20], a device based on a DNA catenane [21], and an origami structure which flipped between states when a linker underwent a transition from B- to Z- form [22]. None of these devices were capable of autonomous operation, and here I present a concept for a synthetic rotary nanomotor which can rotate on its own, and is both constructed from and fuelled by DNA. Inspiration for the structure and mechanism of this motor was provided by the naturally occurring molecular machine ATP synthase [23,24].

**Concept**

The motor consists of an axle and a rotor made using the technique of DNA origami (Fig. 1a) [10-12]. The axle is immobilized on a surface (Fig. 1b) and the rotor revolves around the axle, driven by interactions between the three 'capture units' on the rotor and the single

'binding component' (BC) on the axle (Fig. 2a). In the absence of other strands, the BC is designed to adopt the form shown in Fig. 2b. The BC binds to a capture unit via an unfuelled (Fig. 2c) or fuelled (Fig. 2d) linker (X, Y or Z).

The initial motor state is shown in Fig. 3a. Capture unit 1 is connected to the BC via an unfuelled linker Z and the other capture units have not yet been provided with linkers. Hairpin $a_4$-$a_5$-$a_4$* is opened, and domain $a_5$ is accessible.

Motion is initiated by addition of fuelled linkers X, Y and Z. These complexes bind to capture units and the $a_5$* domain of Y binds to the $a_5$ domain of the BC (Fig. 3b). Toehold-mediated strand displacement ensues, opening the loop of the $\tilde{Z}$ fuel of linker Y. The newly released $a_4$* domain in the $\tilde{Z}$ fuel binds to one of its complements in the fuel loop, with a higher probability of hybridizing to the nearest. This ensures that the subsequent reactions are likely to proceed as intended without interference from this domain.

When the $\tilde{Z}$ fuel loop is opened, the toehold domain $t_z$* is free to bind to its complement $t_z$ on linker Z and $l_z$* can also hybridize to $l_z$ (Fig 3c). The linker Z is displaced from the BC, resulting in hairpin refolding. As strand displacement continues, linker Z is released from capture unit 1, which will be reloaded subsequently by binding of fuelled linker Z from solution. Ultimately, the fuel $\tilde{Z}$ is also stripped away from linker Y.

The conformation of capture units and the BC is now identical to the starting arrangement apart from the identity of the hybridized domains. At this stage the rotor has moved by approximately 1/3 of a turn relative to the axle. Rotation continues as shown in the Supplementary Figures.

Experimental demonstration of the operation of the motor is outside the scope of this paper, but it is instructive to consider what would be involved. Before assembly of the entire motor, polyacrylamide gel electrophoresis should be used to confirm the structure and properties of the strands comprising the BC, capture units, linkers and fuels. It should be verified that the components hybridize as designed, the secondary structure motifs fold correctly and there are no undesired interactions. Formation of the axle and rotor should be confirmed using negative stain transmission electron microscopy or cryo-electron microscopy. Finally, rotation of surface-immobilized motors could be observed using super-resolution fluorescence microscopy or high-speed atomic force microscopy (AFM).

**Conclusion**

An autonomous rotary motor of the type described in this paper could potentially play a role in a molecular computer [25], in combination with DNA logic gates or similar systems [26,27]. Alternatively, it could act as a nanoscale winch, capable of being employed to twist a polymer, or it could be combined with other technologies for use as the 'gatekeeper' for a nanopore. The motor might also have potential for the construction of a modulated nanoscale plasmonic antenna – if a fluorophore was attached to the rotor such that it regularly passed through a gap between a pair of gold nanoparticles, the fluorescence intensity would be considerably enhanced at periodic intervals, potentially by a factor of 100 [28].

As in the case of other DNA devices, the rotary motor presented here is driven by strand displacement reactions. Here, the reactions are co-ordinated to enable continuous autonomous rotation, and the resulting rotary nanomachine has several possible applications, as described above. The design was inspired by ATP synthase and has the potential to extend the capabilities of DNA motors.

**Acknowledgements**

My doctoral research was funded by EPSRC, and resulted in a thesis entitled 'DNA origami assembly'. Incidental to this research was my idea for an autonomous rotary DNA motor, and I developed the theory sufficiently for it to be included as an appendix to my thesis. I am grateful to Dr Beth Bromley for her insightful comments on that early design.

This paper describes a later design for a synthetic autonomous rotary nanomotor made from and fuelled by DNA. The invention is the subject of an International Patent Application under PCT, supported by Isis Innovation Ltd.

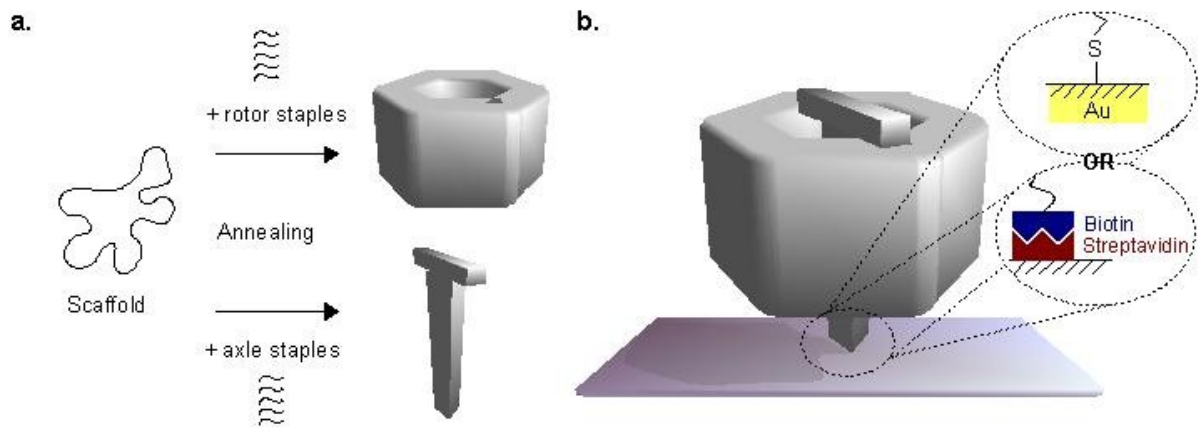

**Figure 1**

**(a)** Assembly of the axle and the rotor using the technique of DNA origami. A long single-stranded DNA scaffold is folded into the desired structure by annealing with a set of 'staple strands' which pin the scaffold together. One staple set is used for the rotor, and a second set for the axle. **(b)** The axle and rotor attached to a surface. The axle could be immobilized either via thiol-gold bonds or biotin-streptavidin interactions.

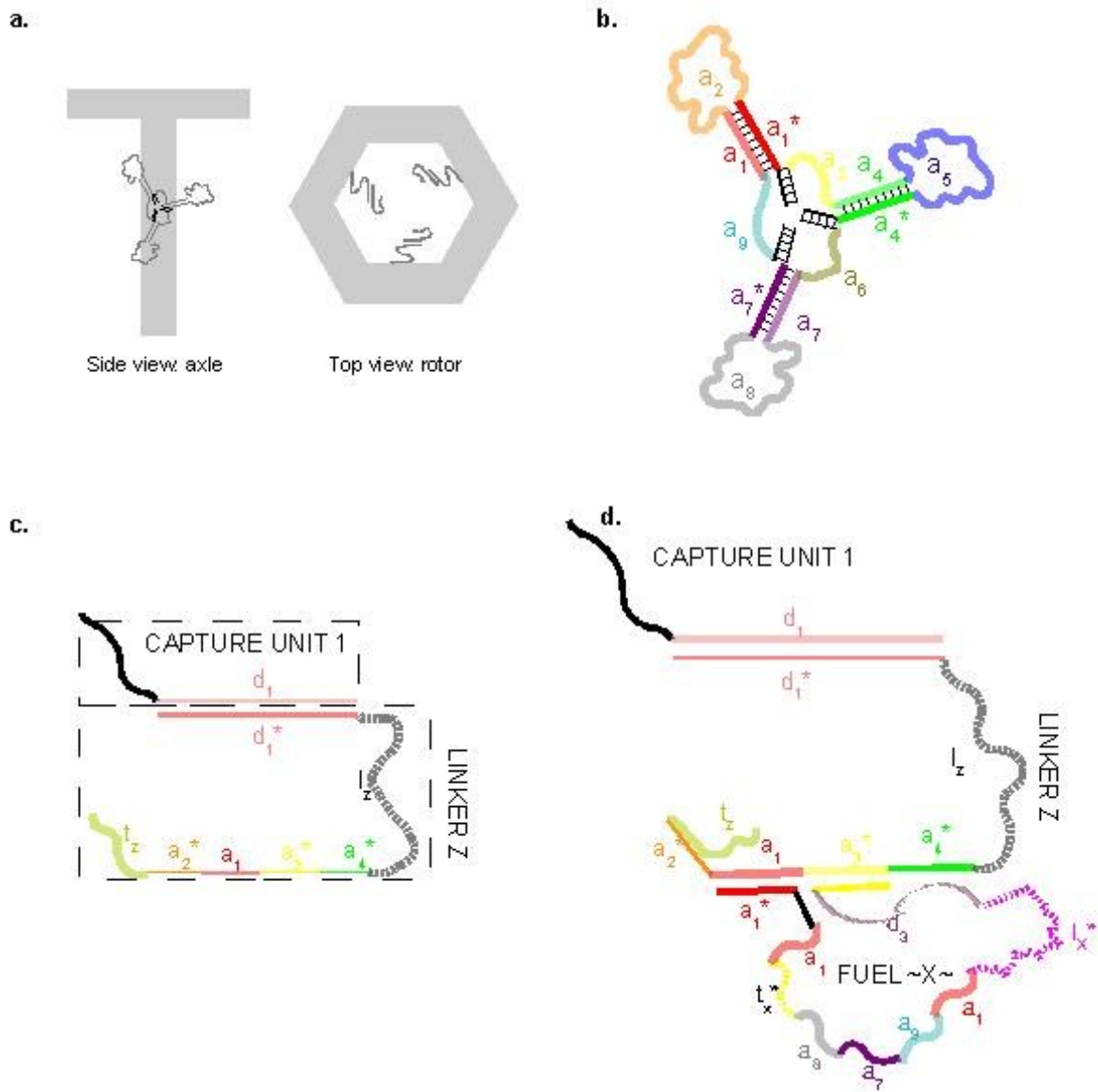

**Figure 2**

**(a)** Side view of axle, showing binding component, and top view of rotor, showing capture units with unfuelled linkers. **(b)** Schematic diagram of binding component, showing the structure adopted in absence of other strands. Domains are colour-coded by sequence identity. Black domains attach the binding component to the axle as shown in (a). **(c)** A capture unit, with unfuelled linker attached. **(d)** The same capture unit and linker, with the fuel bound. In (b) base pairs are shown explicitly; in parts (c) and (d) domains of the same colour are to be taken as hybridized when adjacent to each other. Each 'a' domain is short, such that the a-a* duplex is only just stable at room temperature.

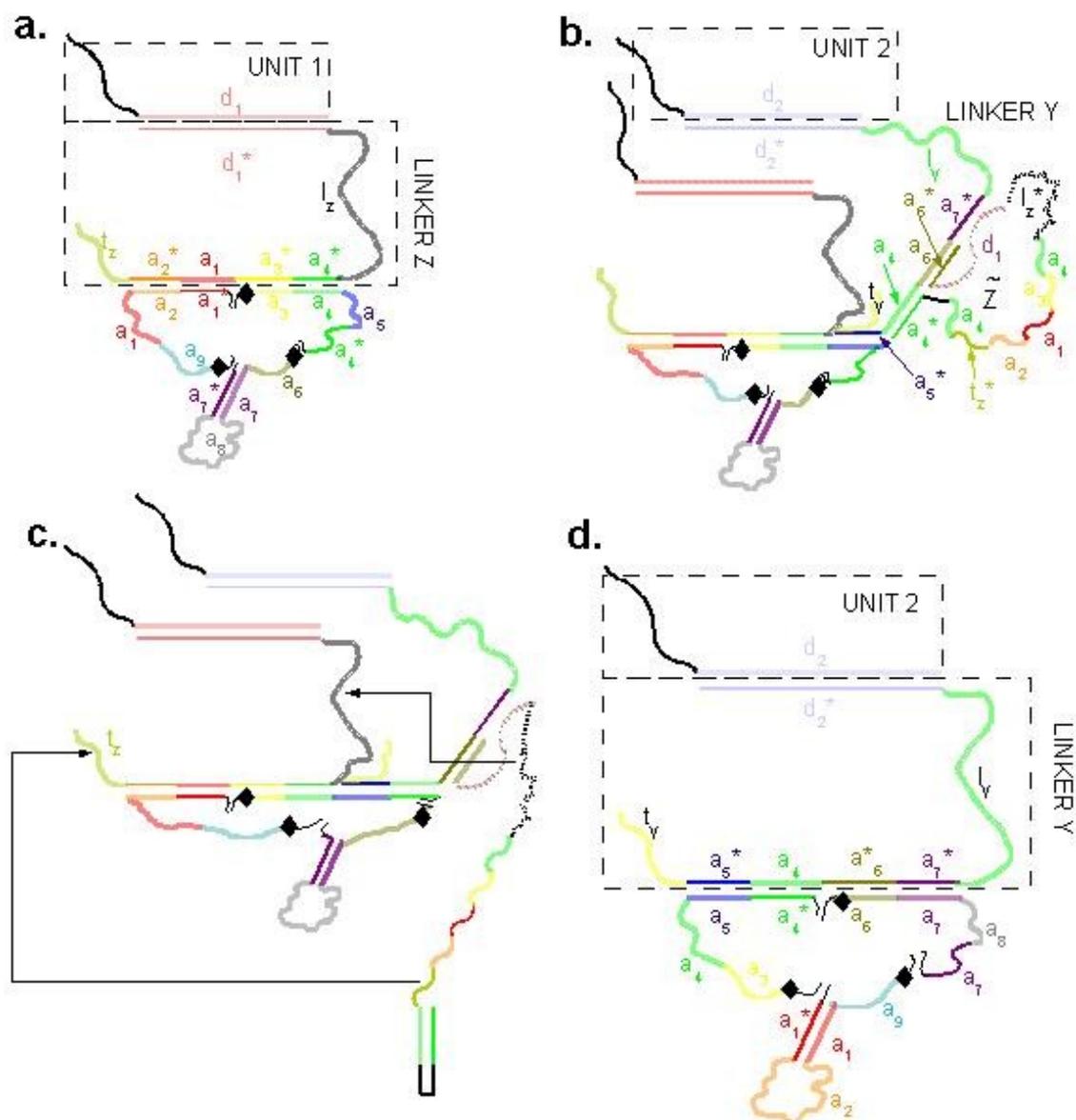

**Figure 3**

**(a)** Initial state of the motor. Capture unit 1 is bound to the binding component via an unfuelled linker Z. Black diamonds mark locations in which the sequence of the binding component is not perfectly matched to that of the linkers; these mismatches (2 or more) slow down displacement processes at critical points of the rotation, with the intention of enhancing the efficiency of the motor. **(b)** The $a_5^*$ domain of linker Y (attached to capture unit 2) binds to the $a_5$ domain of the binding component. Toehold mediated strand displacement ensures, opening the loop of the fuel $\tilde{Z}$. The loop is comparatively large, and further secondary structure may be needed here to suppress unwanted 'leak' reactions. **(c)** The fuel $\tilde{Z}$ displaces the linker Z from both the binding component and the capture unit. Black arrows indicate initiation of toehold-mediated strand displacement. **(d)** The motor has now rotated by 1/3 of a turn; linker Y is now fully bound to the binding component.

**Supplementary Figures**

**Supplementary Figure 1**

State entered by motor after state shown in Figure 3d. Capture unit 3 binds to the binding component via linker X, which is fuelled with fuel $\tilde{Y}$. The next state is shown in Supplementary Figure 2.

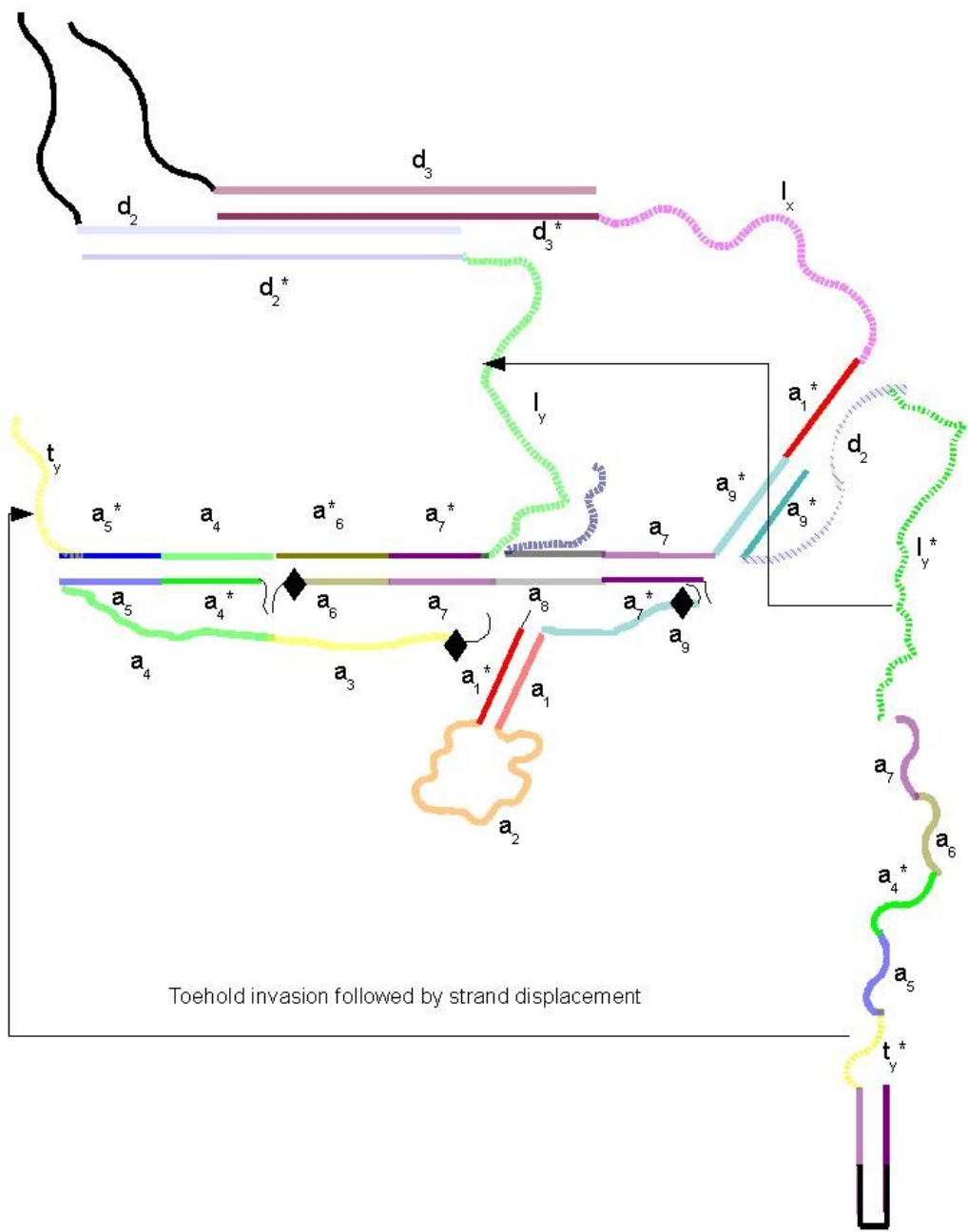

**Supplementary Figure 2**

As the loop in fuel $\tilde{Y}$ is opened, the domains $t_y^*$ and $l_y^*$ are made available for binding. This enables toehold invasion to occur as indicated. Displacement reactions drive the transition to the state shown in Supplementary Figure 3.

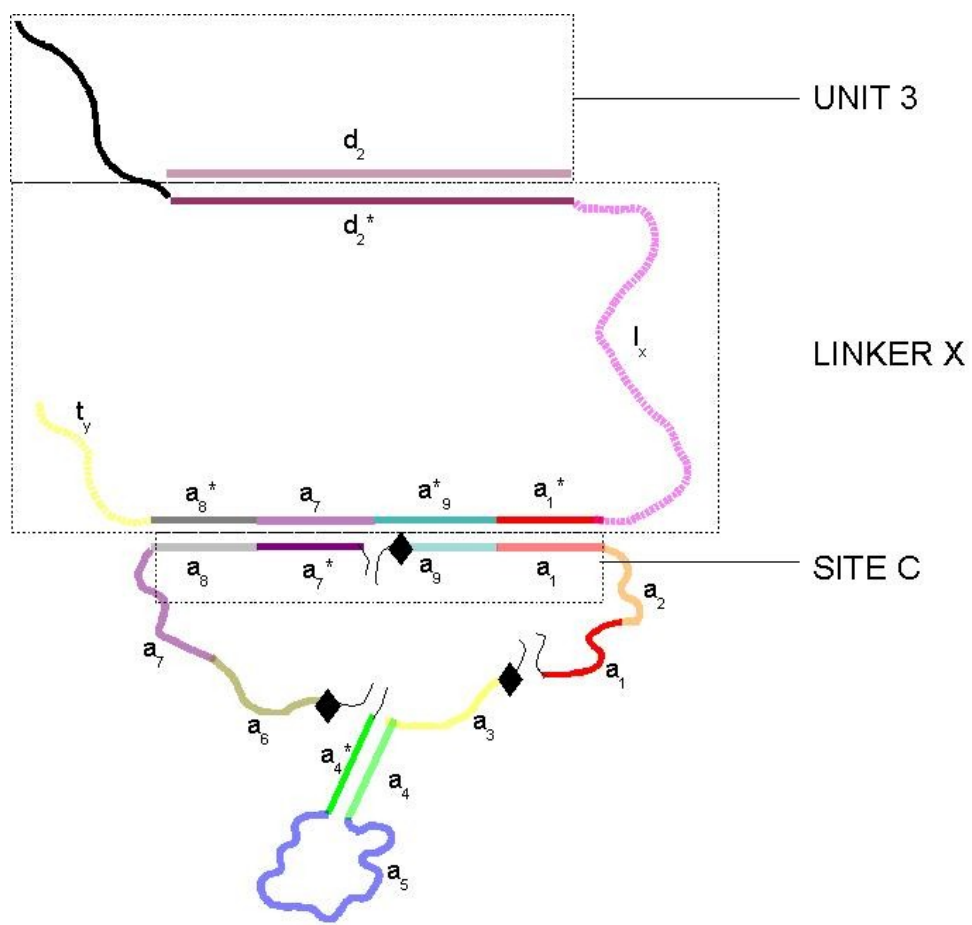

**Supplementary Figure 3**

Now capture unit 3 is bound to the binding component via linker X. Capture unit 2 has been displaced (and will subsequently be supplied with a new fuelled linker from solution). The next state is shown in Supplementary Figure 4.

**Supplementary Figure 4**

The opening of the hairpin $a_1$-$a_2$-$a_1$* makes it possible for capture unit 1 to attach itself to the binding component again, via linker Z. The motor then progresses to the configuration shown in Supplementary Figure 5.

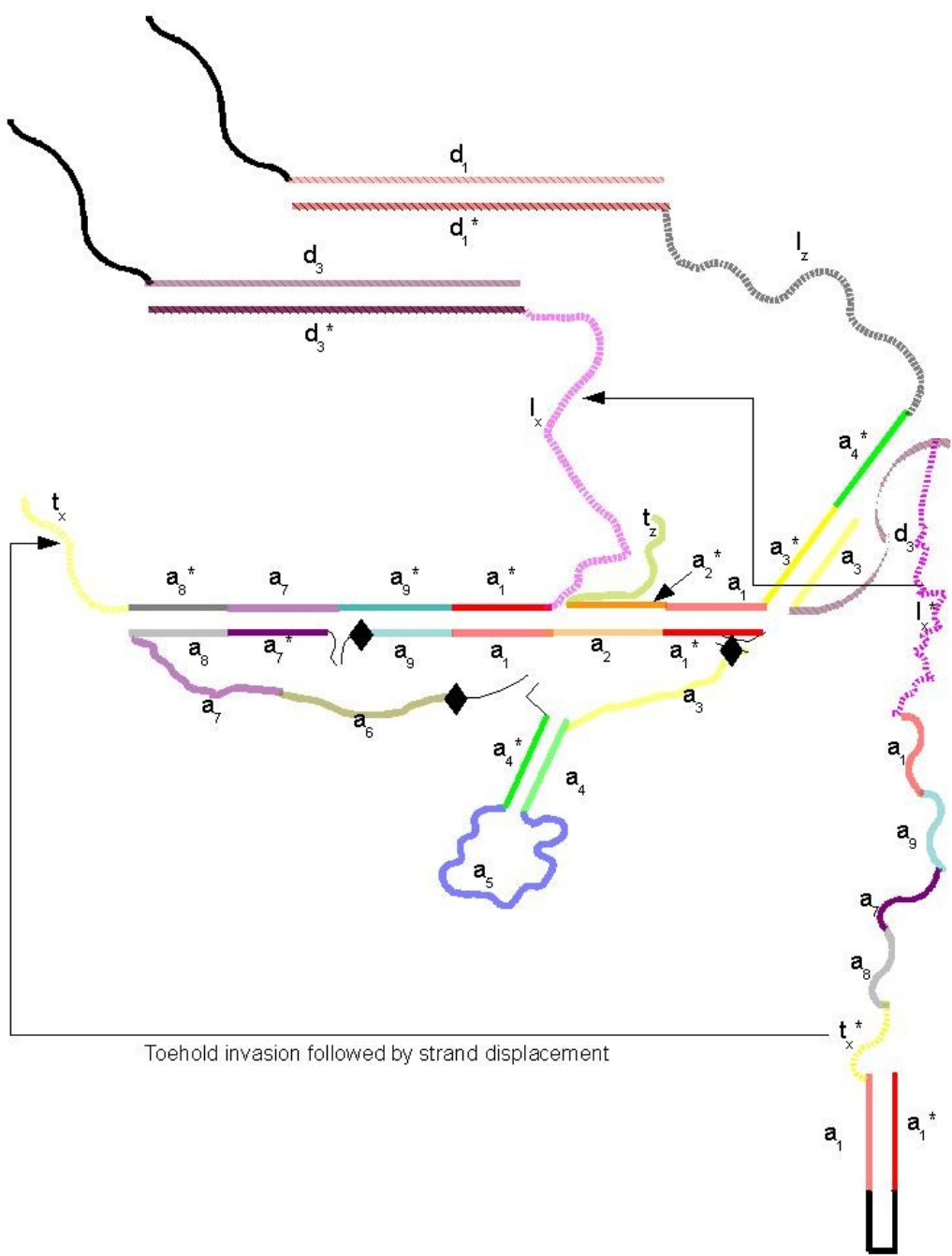

**Supplementary Figure 5**

Opening of the loop in the fuel once again enables toehold invasion and strand displacement, which ultimately strips away linker X. This breaks the link between capture unit 3 and the binding component and the initial state (Fig. 3a) is restored.